\documentclass[aps,amsfonts,amsmath,superscriptaddress,%twocolumn,
showpacs,floatfix,nofootinbib,showkeys]{revtex4-1}

\usepackage{marginnote}

\usepackage{slashed}
\usepackage{color, verbatim}
\usepackage{latexsym}
\usepackage{epsfig}
\usepackage{amsmath}
\usepackage{stackrel}

\usepackage{amssymb}
\usepackage{graphicx}
\usepackage{bm}

\usepackage{hyperref}
\usepackage{epstopdf}
\definecolor{myred}{rgb}{0.7, 0, 0}
\definecolor{myblue}{rgb}{0, 0, 0.7}
\definecolor{mygreen}{rgb}{0.04, 0.7, 0.5}
\hypersetup{colorlinks,citecolor=myred,linkcolor=myblue,urlcolor=myblue,linktocpage=true}

\makeatletter

\newcommand{\be}{\begin{equation}}
\newcommand{\ee}{\end{equation}}
\newcommand{\bea}{\begin{eqnarray}}
\newcommand{\eea}{\end{eqnarray}}

\newcommand{\diag}{\operatorname{diag}}

\begin{document}

\title{
Pulsar Timing Array Stochastic Background from light Kaluza-Klein resonances}

\author{Eugenio Meg\'{\i}as}
\email{emegias@ugr.es}
\affiliation{Departamento de F\'{\i}sica At\'omica, Molecular y Nuclear and  Instituto Carlos I de F\'{\i}sica Te\'orica y Computacional, Universidad de Granada, Avenida de Fuente Nueva s/n,  18071 Granada, Spain}

\author{Germano Nardini}
\email{germano.nardini@uis.no}
\affiliation{Faculty of Science and Technology, University of Stavanger, 4036 Stavanger, Norway}

\author{Mariano Quir\'os\,}
\email{quiros@ifae.es}
\affiliation{Institut de F\'{\i}sica d'Altes Energies (IFAE), The Barcelona Institute of  Science and Technology (BIST), Campus UAB, 08193 Bellaterra (Barcelona) Spain}

\date{\today}

\vspace{2cm}

\begin{abstract} 
\vspace{1cm}
\noindent
 We investigate the potential of the warped-extradimension framework as an explanation for the recently observed stochastic gravitational background at nHz frequencies in pulsar timing arrays (PTA). Our analysis reveals that the PTA data can be effectively accommodated by a first-order phase transition triggered by a radion at the MeV-GeV scale feebly coupled to the Standard Model. Remarkably, this outcome remains robust irrespective of the specific details of the warped extradimension embedding, providing a foundation for future investigations aiming to develop concrete extradimension descriptions of Nature. We also demonstrate that many existing embeddings are not viable, as their radion and graviton phenomenology clash with a MeV-GeV scale radion. As a possible way-out, we sketch a promising solution involving multiple branes, wherein the light radion, graviton, and ensuing light resonances remain consistent with collider bounds and gravity tests.

\end{abstract}

%\pacs{}

%\keywords{}

\maketitle

\section{Introduction}
\noindent
A stochastic gravitational wave background (SGWB) signal can originate from the superposition of numerous independent gravitational wave events, often of astrophysical nature, that are too weak to be individually resolved. Alternatively, the SGWB can arise from intrinsically non-local and stochastic sources which may happen in the primordial universe. The detection of a SGWB in the nHz frequency band is a primary goal of pulsar timing array (PTA) experiments. In this frequency range, astrophysics predicts a SGWB signal primarily generated by inspiraling supermassive back hole binaries (SMBHBs)~\cite{Rajagopal:1994zj, Jaffe:2002rt, Sesana:2008mz}. According to current understanding of galaxy formation and evolution, the SMBHB population produces a SGWB with a strain following a power law characterized by a power index $\gamma \simeq 13/3$~\cite{Phinney:2001di} and an amplitude $A \sim 10^{-15}$ at  frequency of $f=1$ year$^{-1}$~\cite{NANOGrav:2023pdq,Antoniadis:2023aac}. Large deviations from this prediction in the detected SGWB would indicate the presence of misunderstood astrophysics, or the  breaking of the standard model of cosmology and particle physics.

A few years ago, the PTA collaborations released their analyses of the data collected over the past decade or so~\cite{Antoniadis:2022pcn, NANOGrav:2020bcs, Chen:2021rqp, Goncharov:2021oub}. These analyses independently revealed the presence of a \textit{common red noise} contamination in the pulsar arrival timings.   Common red noise is a clear signature of a SGWB signal if accompanied by evidence in Hellings and Downs correlation observable~\cite{Allen:2023kib}. The latter was absent in all sets of data,
maybe due to insufficient statistics, or overlooked systematics, in the analyses~\cite{Allen:2022dzg}. Furthermore, the SGWB that could have explained the observed common red noise would have poorly matched the power-law-like signal produced by SMBHBs for both, the power index~\cite{Antoniadis:2022pcn, NANOGrav:2020bcs, Chen:2021rqp, Goncharov:2021oub}, and possibly the amplitude~\cite{Izquierdo-Villalba:2021prf}.

With the additional PTA data collected in the past couple of years, significant progress has been made. For the first time, there is strong evidence of a detected (isotropic) SGWB~\cite{NANOGrav:2023gor, Antoniadis:2023ott, Reardon:2023gzh, Xu:2023wog}.
 The common red noise has spectral shape compatible to the one reported in the previous data release, and Hellings and Downs correlation is now present~\cite{NANOGrav:2023gor, Antoniadis:2023ott, Reardon:2023gzh, Xu:2023wog}. Intriguingly, the observed SGWB signal can be explained in terms of SMBHB population models only by stretching the population parameters slightly beyond what was previously considered reasonable~\cite{NANOGrav:2023pdq, Antoniadis:2023xlr}, and the hypothesis of a SGWB sourced by first-order phase transitions (FOPTs), on the top or without the expected SMBHB SGWB, is more compatible with the NANOGrav data than the SMBHB-only hypothesis~\cite{NANOGrav:2023hvm}.  

More specifically, a FOPT with gravitational waves primarily triggered by bubble collisions, strength $\alpha_*\gtrsim 1$, reheating temperature $10^{-2}\lesssim T_*/{\rm GeV}\lesssim 10^{1}$, and inverse time duration $\beta/H_*\lesssim 50$ is one of the source candidates yielding the highest Bayes factor in the NANOGrav analysis~\cite{NANOGrav:2023hvm}. However, in qualitatively similar ranges of values of $\alpha_*$, $\beta/H_*$ and $T_*$, also the FOPT SGWB generated by the sound-waves and turbulence contributions fits well the PTA observations~\cite{NANOGrav:2023hvm, Antoniadis:2023xlr}.

Shortly after the PTA data release of a few years ago, various studies explored the possibility of explaining the observed red noise as a result of a cosmological FOPT~\cite{Kobakhidze:2017mru,Arunasalam:2017ajm,NANOGrav:2021flc, Neronov:2020qrl, Ratzinger:2020koh, Nakai:2020oit, Chiang:2020aui, Li:2021qer, Banerjee:2021oeu, Xue:2021gyq, Brandenburg:2021tmp, Borah:2021ftr, Xue:2021gyq, Freese:2022qrl, Bringmann:2023opz,Addazi:2023jvg} (for a recent review see e.g.~\cite{Athron:2023xlk})~\footnote{For other possible cosmological explanations see e.g.~Refs.~\cite{Ashoorioon:2022raz,King:2023cgv,Madge:2023cak}. A statistical comparison of them based on the calculation of the Bayes factor can be found in Ref.~\cite{NANOGrav:2023hvm}.}. The conclusions drawn from these studies are expected to remain valid even with these data update, as a few extra years of data unlikely could significantly alter the common red noise observed over more than a decade. However, with the evidence of the SGWB reaching 4-$\sigma$, it is timely to take a step forward. Indeed, despite the anticipation that PTA experiments were likely on the cusp of strengthening the previous results favoring new physics, explicit particle physics frameworks featuring the required FOPT have been scarce.~\footnote{In first approximation, the concrete models in the literature featuring a FOPT  successfully fitting the common red noise reported a couple of years ago  (see e.g.~Refs.~\cite{Borah:2021ftr, Li:2021qer, Lerambert-Potin:2021ohy, Banerjee:2021oeu, Chiang:2020aui}), essentially rely on: invoking FOPTs in secluded sectors which are poorly  constrained as hidden to the SM particle interactions; attempting to move the QCD phase transition from the crossover regime to the FOPT one; embedding the SM in setups where the GeV-MeV new physics compatible with current experiments comes at the expenses of high tuning or cumbersome extensions. For Bayesian analyses of different models describing the signal, see e.g.~Refs.~\cite{NANOGrav:2023hvm,Bian:2023dnv}.} This scarcity highlights the challenges faced by extensions of the Standard Model with FOPTs at the MeV-GeV scale in overcoming existing phenomenological constraints.

In this article, we investigate the feasibility of a BSM setup that we find particularly intriguing. We introduce a simple, yet versatile, warped extradimensional model adapted to the GeV scale. This model predicts unavoidable FOPTs and offers a phenomenology that, at present, remains comfortably consistent with, albeit not invisible to, current or forthcoming experimental searches. While more complex variations can be developed without compromising its fundamental characteristics, our proposed model serves as a promising foundation for investigating the implications of warped extradimensions in the GeV regime relevant for the PTA SGWB discovery.

Our analysis is structured as follows. In Sec.~\ref{sec:model} we introduce the setup and summarize its signatures at some observables. In Sec.~\ref{sec:FOPT} we prove that the model can easily induce a FOPT compatible with the SGWB discovery announced by the PTA Collaborations and, in the parameter region fitting the PTA data, its main phenomenological signatures are broadly compatible with current constraints. In Sec.~\ref{sec:embeddings} we study possible embedding options of the SM fields in the model, as well as their possible phenomenological signatures due to the interactions of these fields with the graviton and radion sectors of the model. We finally devote Sec.~\ref{sec:conclusions}  to our conclusions.
%--------------------------------------------------------------------------------------

\section{The model}
\label{sec:model}
\noindent
The candidate model that we propose as an explanation of the FOPT possibly at the origin of the PTA SGWB background, is a slight variation of the Randall-Sundrum (RS) model. The RS model has been the starting point for numerous studies, providing a natural solution of the hierarchy problem, a natural strong FOPT at the EW scale, and a rich phenomenology at the LHC. Here we review the main ideas of the model, and while skipping the intermediate steps of calculations known in the literature, we summarize some key results and adapt them to the scenario we are interested in. In this brief overview we follow Refs.~\cite{Megias:2018sxv, Megias:2020vek,Girmohanta:2023sjv} and references therein.

\subsection{The model at zero temperature}
\label{sec:zeroT}
\noindent
The Randall-Sundrum (RS) model is a 5D setup built on Anti-de Sitter (AdS) spacetime with line element~\footnote{The original RS model assumes $A(r)\propto r$ where $r$ is the extra dimension in proper coordinates. Here we consider a more generic function $A(r)$ that still warps the extradimension while fixing the branes distance and thus the radion mass. Depending on the specific profile of $A(r)$, the technical literature may call the model ``warped", ``soft-wall", or other names.  With an abuse of language, in this article we denote this wide class of models simply as RS or warped.}
\begin{equation}
ds^2_{{\rm RS},T=0} = g_{MN}dx^M dx^N\equiv e^{-2A(r)} \eta_{\mu\nu} dx^\mu dx^\nu-dr^2\,,    \qquad \textrm{with} \qquad \eta_{\mu\nu}=\diag(1,-1,-1,-1) \;,
\quad d A(r)/dr > 0 \;,
\label{eq:line_elementAdS}
\end{equation}
with Latin (Greek) indices  running over the five (four) spacetime dimensions. Along the $r$ direction, $n$ Minkowskian 4D spaces $\mathcal{B}_a$, dubbed branes, are located at  $r=r_a$, with $a=0,I_1,\dots,I_m,1$, where $\mathcal B_0$ is the UV (Planck) brane, $\mathcal B_1$ is the IR brane, and we allow in principle a number $m$ of hypothetical intermediate branes $\mathcal B_{I_i}$ ($i=1,\dots,m$), at intermediate scales between the Planck scale and the scale of the $\mathcal B_1$ brane, which will be considered to be at the MeV-GeV scale. The total number of branes would then be $n=2+m$. 
Due to the warped factor $A(r)$, the energy scale involved in the brane $\mathcal{B}_{a_2}$ is exponentially suppressed with respect to the one of the brane $\mathcal{B}_{a_1}$, if $a_1<a_2$, with the suppression depending on the distance $r_{a_2}-r_{a_1}$. The RS setup thus provides an elegant solution for scenarios featuring hierarchical energy scales. In this paper we will consider two possible scenarios: a model with $m=0$ (no intermediate brane), just like the RS scenario, and a model with $m=1$, i.e.~with an intermediate brane at the TeV scale, but of course more general scenarios can be considered for other purposes.

The AdS metric exhibits conformal invariance along $r$. The localization of the branes requires a stabilization mechanism that breaks this symmetry. The Goldberg-Wise mechanism is a well-known realization of this breaking~\cite{Goldberger:1999uk}. 
%  is bound by two Minkowskian 4D spaces, dubbed UV and IR branes, respectively located at $r=r_0=0$ and $r=r_1>r_0$.   
  It involves a (extremely heavy) scalar field, $\phi$, that can propagate in the bulk (i.e.~the space between the branes) and has bulk potential $V(\phi)$   
and brane potentials $\Lambda_a(\phi)=\Lambda_a+\frac{1}{2}\gamma_a(\phi-v_a)^2$, with $\Lambda_a$, $v_a$ and $\gamma_a (\gg 1$ in the stiff limit) some positive constants (see Refs.~\cite{Goldberger:1999uk, Megias:2018sxv} for details). In this case, the five-dimensional action of the model reads as
%
%\begin{align}
%S &= \int d^5x \sqrt{|\det g_{MN}|} \bigg[ 
%%
%-\frac{1}{2\kappa^2} R +
%\frac{1}{2} g^{MN}(\partial_M \phi)(\partial_N \phi) \nonumber \\
%&-V(\phi) + \mathcal{L_{\rm bulk}}\bigg] - \sum_{a} \int_{\mathcal{B}_a} d^4x \sqrt{|\det \bar g_{\mu\nu}|} (\Lambda_a(\phi) + \mathcal{L}_{a}+ S_{\rm GHY}) \,, \label{eq:action}
%\end{align}
\begin{align}
S &= \int d^5x \sqrt{|\det g_{MN}|} \bigg[ 
-\frac{1}{2\kappa^2} R +
\frac{1}{2} g^{MN}(\partial_M \phi)(\partial_N \phi) -V(\phi) + \mathcal{L_{\rm bulk}}\bigg] \nonumber  \\
&\qquad\qquad- \sum_{a} \int_{\mathcal{B}_a} d^4x \sqrt{|\det \bar g_{\mu\nu}|} (\Lambda_a(\phi) + \mathcal{L}_{a}+ S_{\rm GHY}) \,, \label{eq:action}
\end{align}
where $S_{\rm GHY}$ is the Gibbons-Hawking-York term canceling out the boundary terms of the 5D geometry variations, the induced metric on the branes being $\bar g_{\mu\nu} = e^{-2 A(r)} \eta_{\mu\nu}$, and finally $\kappa^2\equiv 1/(2 M_5^3)$ where $M_5$ is the 5D Planck scale, a parameter of order the 4D Planck scale $M_{\rm Pl}$. At this point, one can see that the potentials of $\phi$ induce some potentials for $r_a$. In the AdS/CFT correspondence picture, this corresponds to the condensation  of the $SU(N_a)$ (conformal) gauge fields of the $\mathcal{B}_a$ brane.

The values of $r_a$ of the potentials can be calculated by solving the equations of motion (EoM) for $\phi$ and integrating out the heavy degrees of freedom~\cite{Goldberger:1999uk}. Hereafter, we can limit ourselves to the phenomenology of the two most IR branes with a coordinate difference $\Delta r$~\footnote{In our convention, for the case of just two branes ($n=2$), $\mathcal B_0$ is the usual UV (Planck) brane which coincides with the UV boundary of the 5D space and $\Delta r=r_1-r_0$. For the case of three branes ($n=3$), $\mathcal B_0$ is still the UV brane, $\mathcal B_I$ is an intermediate brane (which can be at the TeV scale) while $\mathcal B_1$ is the IR boundary, a dark brane with the lowest scale in the theory. In this case $\Delta r=r_1-r_I$.}. They indeed involve, respectively, the next-to-lowest and the lowest energy scales of the whole picture, and the study of their phenomenology suffices for the purpose of our analysis, namely the FOPT interpretation of the PTA SGWB.~\footnote{We could generalize our calculations  to the case of two generic adjacent intermediate branes $\mathcal{B}_{I_i}$ and $\mathcal{B}_{I_{i+1}}$, but the essence of our analysis would not change.} As only the difference $\Delta r$ matters, we fix the lowest coordinate by convention.~\footnote{In the case of two branes, $r_0$ is the location of the UV brane, which can be freely fixed as it is an integration constant. In the case of three branes $r_I$ has been stabilized by a previous, highest temperature, phase transition that might be detected by LISA and/or ET.} To determine the potentials of their positions, we solve the EoMs with the ``superpotential" method~\cite{Gubser:1999vj}. This method enables us to reliably explore the parameter region leading to sizeable backreaction on the metric~\cite{Konstandin:2010cd}. (In the region with small backreaction, the cosmology of the model tends to get stuck in the wrong phase~\cite{Creminelli:2001th, Girmohanta:2023sjv}; see discussion later on).
By introducing an expansion parameter $u\ll 1$~\cite{Megias:2020vek}, we obtain that  $r_1$  develops the potential~\cite{Goldberger:1999uk, Megias:2020vek}
\be
\bar U_{1,\rm eff}(\bar r_1)=2 u^2 \bar v_1^2 (\bar r_1^0-\bar r_1)
\left[e^{4A_0(\bar r_1^0)-4A_0(\bar r_1)}-1\right]\, e^{-4 A_0(\bar r_1)} 
\,%
+6\lambda_1 e^{-4 A_0(\bar r_1)} \,,  \label{eq:Uefftuned_approx}
\ee
where $\bar r_1^0$ is defined by the condition
$v_1\equiv v_0e^{u \bar r_1^0}$,
 $A_0(r)$ is the leading order approximation of $A(r)$ at $r>0$, and finally $\bar v_a\equiv \kappa v_a$ and $\bar r_1\equiv r_1/\ell$. Notice that in Eq.~\eqref{eq:Uefftuned_approx}, $\lambda_1$ is a free negative parameter whose absolute value is in the range 
 $\mathcal{O}(0.1 - 10)$. The position of the minimum of the potential $\bar U_{1,\rm eff}^0$ sets the value $r_1^{\rm min}$ at which the  brane $\mathcal{B}_1$ is stabilized.
 
Alternatively, one can reparametrize the degree of freedom associated to the $r_1$ coordinate in terms of the radion scalar field $\chi_1$ defined as
\be
\chi_1\equiv\bar\chi_1(r_1)/\ell= \ell^{-1} e^{-A_0(r_1)}\,.
\label{eq:chi}
\ee
For consistency, its potential $V_{\rm r_1}$ will have minimum at   $\rho \equiv e^{-A_0(\bar r_1^m)}/\ell$. Moreover, after some algebraic manipulation of Eq.~\eqref{eq:Uefftuned_approx}, one finds 
\be
V_{\rm r_1}(\chi_1)= \frac{N^2\rho^4}{8\pi^2} e^{4 A_0(\bar r_1^m)}\bar U_{1,\rm eff}(\bar r_1(\bar \chi_1)) \,,
\label{eq:Vrad}
\ee 
where the function $\bar r_1(\bar \chi_1)$ is the inverse of Eq.~\eqref{eq:chi} and $N = (2\ell)^{3/2}\pi/\kappa$ is the number of colors of the $SU(N)$ symmetry of the AdS/CFT correspondence for the next-to-last brane.

The described procedure is not specific of the behaviour of the brane at $r_1$. Should we have been interested in energies processes not below the energy of the next-to-last brane, we could have repeated the procedure for the positions $r_0, r_{I_1}, \dots$ as well, introduced radion fields for each position,  and obtained expressions conceptually similar to Eqs.~\eqref{eq:Uefftuned_approx}--\eqref{eq:Vrad} also for them.

\subsection{The model at finite temperature}
\label{sec:finiteT}
\noindent
As mentioned earlier,  our interest lies in the phenomenology and the FOPT associated with the lowest $\mathcal B_1$ and next-to-lowest $\mathcal B_N$ branes of the complete RS picture~\footnote{Just remind the reader that according to our notation, for $n=2$, $\mathcal B_N=\mathcal B_0$, and for $n=3$, $\mathcal B_N=\mathcal B_I$.}. Correspondingly, we limit ourselves to consider temperatures $T$ below the scale of the brane $\mathcal B_N$. Only the dynamics of the $\mathcal{B}_N$ and $\mathcal{B}_1$ is then relevant, and we now focus on it (see Ref.~\cite{Girmohanta:2023sjv} for the finite-temperature picture in the case with several branes).

In quantum field theory at finite temperature, the temperature $T$ replaces the time coordinate and the $x_0$ component of the spacetime gets compactified in a circle.  In this regime, the 5D gravity action leading to the AdS solution in Eq.~\eqref{eq:line_elementAdS} now admits two geometric/gravitational solutions~\cite{Witten:1998zw,Creminelli:2001th}: the previous AdS solution of Eq.~\eqref{eq:line_elementAdS} where the $x_0$ direction is compactified, say $ds^2_{{\rm RS,}T}$ its line element, and the so called Anti de Sitter-Schwarzschild (AdS-S) solution. The latter corresponds to the metric of black hole with line element
\begin{equation}
ds^2_{\rm AdS-S}= - h(r)^{-1} dr^2+e^{-2A(r)}[h(r)dt^2-d\vec x^2]\,,
\end{equation}
with the event-horizon singularity at  $r=r_h>0$. Notice that both solutions are equivalent at $r\to \infty$. The existence of the brane of next-to-lowest energy scale, $\mathcal{B}_N$ (with $r_N$ stabilized), is then allowed in both cases. Instead, the presence of brane $\mathcal{B}_1$ at $r=r_1$ is permitted in both solutions only when $r_h>r_1$. Indeed, roughly speaking, for $r_h < r_1$, in the physical space containing the brane $\mathcal{B}_N$, the event horizon masks the brane $\mathcal{B}_1$. Then, when the zero-temperature RS action is heated up to the AdS-S background, the relevant 5D  action is the one in Eq.~\eqref{eq:action} with the $\mathcal B_1$ terms omitted. The degrees of freedom in the AdS-S and RS$_{T}$ phases are then different: in the former, the light fields localized in $\mathcal{L}_1$ disappear while the $SU(N)$ symmetry is restored; in the latter, the $SU(N)$ fields condensate and the fields of the lightest energy scale emerge. Due to this feature the AdS-S and RS$_{T}$  phases are also respectively dubbed as ``deconfined" and ``confined" phases.

Similarly to what happens to $r_1$ in the zero-temperature case, the gravity action itself does not specify the value of $r_h$; it is the backreaction of the non-gravity content that stabilizes it. The metric solution with $r_h$ at its minimum $r_h^{\rm min}$ is the ``on-shell" (or classical) 
gravitational solution. This implies that, if the scalar field $T_h$ takes the role of the degree of freedom along the direction $r_h$ (and hence the  potential of $r_h$ is translated into a potential of $T_h$), at the minimum of the $T_h$ potential we must find $\langle T_h \rangle = T$, with $T$ being the Hawking temperature of the black hole. In analogy with the radion, we define~\footnote{We remind that $T$ is assumed to be below the energy scale of $\mathcal{B}_{a}$ controlled by the distance $r_{a}-r_0$ (where $r_a$ is the position of any other brane whose energy scale is known). This is consistent with the described picture taking $r_h>r_0$ as $T$ is also the position of the potential minimum of $T_h$ and, due to  Eq.~\eqref{eq:Th}, sets the scale of $r_h$.}
\be
T_h\equiv \exp[-A_0(r_h)] /(\pi \ell)
\label{eq:Th}\;.
\ee

To determine the position at which $r_h$ stabilizes, we follow Ref.~\cite{Megias:2020vek}. As previously done for the zero temperature case, we introduce a small expansion parameter $u$ and use the  ``superpotential" method~\cite{Gubser:1999vj} to solve the EoM. In the solution we take into account that the blackening factor  $h(r)$ must satisfy the boundary and regularity conditions $h(0)=1$ and $h(r_h)=0$. This leads to
\be
h(r)\simeq 1-e^{4[A_0(r)-A_0(r_h)]} \qquad \textrm{and} \qquad \frac{dA_0(r)}{d \bar r}=1+\mathcal  O(u)\,, 
\ee
from which, after multiple algebraic manipulations, one can derive the finite-temperature effective potential, or equivalently, the free energy of $T_h$. Here we report on its value at the minimum, which is located at $T_h=T$ as expected (see Ref.~\cite{Megias:2020vek} for the full expression):
\be
F_{\rm min}^{\rm AdS-S}(T)\simeq -\frac{\pi^4\ell^3}{\kappa^2}T^4 = - \frac{\pi^2}{8} N^2 T^4  \;.
\label{eq:Fmin}
\ee

The quantity $F_{\rm min}^{\rm BH}(T)$ is the ``geometric contribution" (i.e.~due to the singularity) to the total free energy of the model in the AdS-S phase~\cite{Creminelli:2001th}. No similar contribution arises in the free energy of the AdS phase. However, a further contribution comes from the plasma made of the field content of the two phases. Accounting for them leads to~\cite{Nardini:2007me}
\begin{align}
F_d(T)&=E_0+F_{\rm min}^{\rm AdS-S}(T)-\frac{\pi^2}{90}g_d^{\rm eff}T^4 \,, \nonumber\\
F_c(T)&=-\frac{\pi^2}{90}g_c^{\rm eff}T^4 \,,
\label{eq:free}
\end{align}
where $g_{d/c}^{\rm eff}$ is the number of relativistic degrees of freedom in the deconfined/confined phase, and $E_0=V_{\rm r_1}(0)-V_{\rm r_1}(\rho)$ is the potential gap between the two phases in the limit $T \to 0$. The main quantities controlling the free energies are then: the value $N$ of the $SU(N)$ conformal group entering in $F_{\rm min}^{\rm AdS-S}(T)$, the radion potential parameters $v_0$, $v_1$, $\lambda_1, \rho$ modulating $E_0$, and the field content of $\mathcal{L}_{\rm bulk}$, $\mathcal{L}_0$ and $\mathcal{L}_1$ which sets $g_{d}^{\rm eff}$ and $g_{c}^{\rm eff}$. At this point one may already correctly guess that the details of the brane and bulk Lagrangians will play a minor role in our final conclusions because the field content only appears in the number of relativistic species.

\subsection{The phase transition picture}
\label{sec:picture}
\noindent
As just explained, the model in Eq.~\eqref{eq:action} has two competitive phases at temperatures below the next-to-last energy scale of the model, or equivalently for $r_h>r_N$. Both phases have the brane $\mathcal{B}_N$ at $r=r_N$, but the brane $\mathcal{B}_1$ only exists in the AdS phase with the value of $r_1$ already stabilized. It is like that the event horizon at $r_h$ somewhat hides the brane $\mathcal{B}_1$ at $r_1$ when $r_h<r_1$. It is however worth stressing that the two geometries match  at  
$r\to\infty$ as well as at $r_1=r_h\to +\infty$. In the coordinate plane $\{r_h,r_1\}$, 
the deconfined and confined phases can then be described as~\footnote{Here we denote by $r_1^{\rm min}$ the value of $r_1$ at the minimum of the potential, while $r_h^{\rm min}$ is given by the value of $r_h$ at the minimum of the potential as a function of $T_h$ in the AdS-S phase where $T_h^{\rm min}=T$ and thus $r_h^{\rm min}/\ell=-\log(\pi\ell T)$.} $\{r_h^{\rm min}, +\infty \}$ and $\{+\infty, r_1^{\rm min}\}$, and one can move from the former to the latter via the path $\{r_h^{\rm min}, +\infty \} \to \{+\infty, +\infty \} \to \{+\infty, r_1^{\rm min}\}$ along which the metric is kept regular. In terms of the fields $\chi$ and $T_h$, this path is equivalent to $\{T_h  =T, \chi_1 = 0 \} \to  \{T_h  =0, \chi_1 = 0 \}  \to \{ T_h  =0, \chi_1 = \rho \}$. To determine the value of the free energy at the intermediate step, one can notice that the configuration $\{T_h=0, \chi_1 = 0\}$ is an on-shell solution when $T=0$, which implies that
 the free energy at the intermediate step is equal to $F_d(0)$ and is hence larger than the free energies of the final and initial phase. The phase transition between these two phases is then of first order. In particular, the direction of the phase transition occurring along the history of the universe, is the one above described, since $|F_c| \ll |F_d|$ at high enough $T$ (but still fulfilling the condition $r_h>r_N$)~\footnote{If one considered generically high temperatures, the additional phases with $r_{I_i}<r_h<r_{I_{i+1}}$, would enrich the picture. Multiple FOPTs would then be possible~\cite{Banerjee:2021oeu}, but if there is the hierarchy $r_{I_i}\ll r_{I_{i+1}}$, only one would dominate the SGWB at PTA frequencies. This is why we focus only on one of them in the present analysis.}.

A peculiarity of this FOPT is that, potentially, a short epoch of inflation can precede the onset of the bubble nucleation, and the phase transition can end up with a sizable reheating. Indeed, the considered FOPT is often very strong, leading to a sizable supercooling and thus to $T_n\ll T_c$, with $T_c$ implicitly defined by $F_c(T_c)=F_d(T_c)$. In the deconfined phase, inflation starts at the temperature $T_i$, obtained by imposing that the energy density in the deconfined phase,
\be
\rho_d(T)=E_0+3\pi^4\ell^3 T^4/\kappa^2+\pi^2g_d^{\rm eff}T^4/30 \;,
\label{eq:rhod}
\ee
be dominated by the vacuum energy $E_0$. This yields
\begin{equation}
T_i\simeq T_c \left[ 3+4 g_{d}^{\rm eff}(T_n)/15N^2  \right]^{-1/4} \,,
\label{eq:Ti}
\end{equation}
which leads to $ N_e=\log(T_i/T_n)$   e-folds of inflation provided $T_i>T_n$. This entropy injection would dilute all abundances, in particular the baryon-over-entropy ratio which is fixed at the BBN epoch to $\eta_B=n_b/s\sim 10^{-10}$. Assuming that baryogenesis takes place at the electroweak phase transition (electroweak baryogenesis), or earlier (e.g.~leptogenesis), this means that the entropy injection triggered by the FOPT has to be taken into account in the baryogenesis mechanism, and the required value of the generated $\eta_B$ has to compensate for this late entropy production. 

Moreover, the energy balance requires that at the end of the FOPT, the energy density in the deconfined phase, $\rho_d(T_n)$, is converted into radiation density in the confined phase, $\rho_c(T_R)$, which implies~\footnote{The balance is correct up to a small amount of energy  going into GWs. Notice that the transition is extremely fast~\cite{Nardini:2007me}, so that the expansion of the universe is neglected in the energy balance.}
\be
\frac{4}{15N^2} g_c^{\rm eff}T_R^4=T_c^4+\left(3+\frac{4}{15N^2}g_d^{\rm eff}\right) T_n^4 \,,
\label{eq:TR}
\ee
where $T_R$ is the final reheating temperature. 

Overall, these expressions show that $T_R$, $T_c$ and $T_n$ are mildly sensitive to  variations of $g_c^{\rm eff}$ and $g_d^{\rm eff}$ due to the fourth power dependence. This is an extra hint to the fact that our numerical results of the FOPT are almost independent of the particular model setup that specifies $\mathcal{L}_{\rm bulk}$, $\mathcal{L}_0$ and $\mathcal{L}_1$.

\section{Analysis of the phase transition and compatibility with the SGWB PTA data}
\label{sec:FOPT}
\noindent
A FOPT is characterized by the following thermodynamic quantities: the inverse time duration $\beta/H_*$, the strength $\alpha_*$, the nucleation temperature $T_n$. From these, other quantities follow, e.g.~$T_R$ in Eq.~\eqref{eq:TR}. We compute the  thermodynamic quantities by solving numerically the $O(3)$ and $O(4)$ bounce equations along the aforementioned path $\{T_h  =T, \chi_1 = 0 \} \to  \{T_h  =0, \chi_1 = 0 \}  \to \{ T_h  =0, \chi_1 = \rho \}$. The procedure is standard in the FOPT literature and we omit its description (see Ref.~\cite{Megias:2020vek} for the details of the procedure we adopt).

Once nucleated, the FOPT bubbles expand and eventually collide. During this process, they perturb the plasma from its equilibrium. 
Both the bubble collisions and the plasma motion generate a SGWB.  As the plasma motion can occur via turbulence and sound waves, in total three dynamical mechanisms contribute to the final SGWB production:  the ``bubble collision", the ``sound waves" and the ``hydrodynamic turbulence"~\cite{Caprini:2015zlo, Caprini:2019egz}. For each of them, the produced SGWB frequency shape  at the time of production has its own dependence on $\alpha_*$, $\beta/H_*$ and $v_w$, with $v_w$ being the bubble expansion velocity. The temperature $T_R$ enters to take into account the redshift due to the expansion of the Universe between the times of the signal production and detection~\cite{Caprini:2019egz}. 

In their recent analyses, the NANOGrav, EPTA and Indian~PTA experimental collaborations also provide the interpretation 
of the observed SGWB signal in view of a FOPT source. They consider each mechanism once at the time. In principle, these mechanisms are not mutually exclusive so that combinations of them, weighted by some efficiency factors, are possible. This assumption is however reasonable for the PTA analyses. Indeed, the PTA measurements are sensitivity to a rather narrow GW frequency band and, inside it, a single contribution does not dominate over the others only for peculiar FOPT combinations. Likewise, we analyze one SGWB contribution at the time too. 

\begin{figure*}[t]
 \includegraphics[width=1\textwidth]{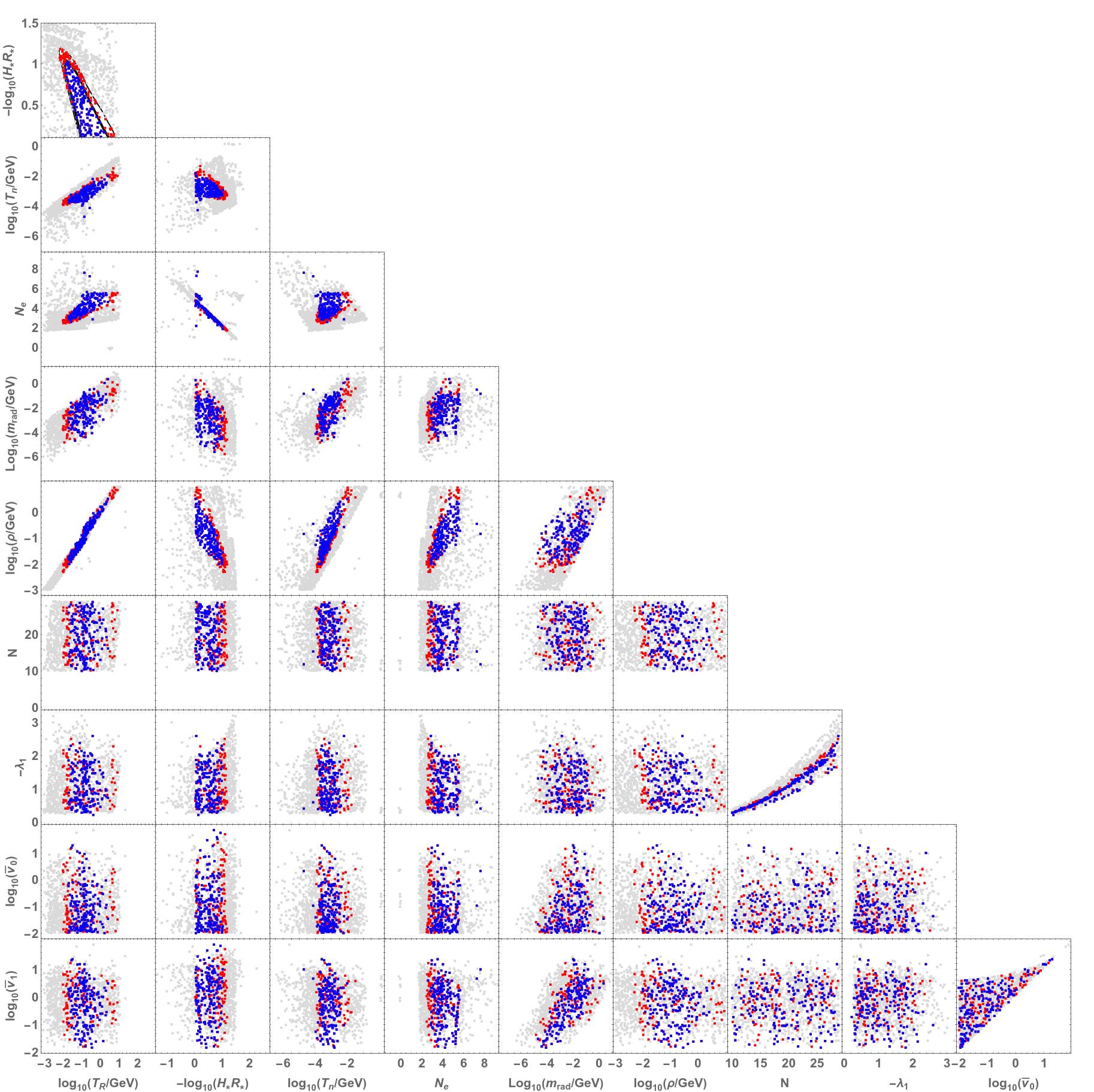} 
\caption{\it Scatter plots of the inputs for the RS framework parameters ($\lambda_1$, $N$, $\rho$) and the resulting FOPT quantities ($T_ R$, $H_* R_*$, $T_n$, $T_i$). Blue points fall in the NANOGrav 95\% favorite region of the ``bubble collision only" hypothesis, the region covered by blue + red points corresponds to the ``bubble collision + SMBHB" hypothesis, and the gray points belong to none of the previous regions. In the top panel, the thin (outer) and thick (inner) lines enclose the NANOGrav 95\% C.L.~favorite regions for the ``bubble collision + SMBHB" hypothesis and  ``bubble collision only" hypothesis, respectively.}
\label{fig:collision}
\end{figure*}

\begin{figure*}[t]
 \includegraphics[width=1\textwidth]{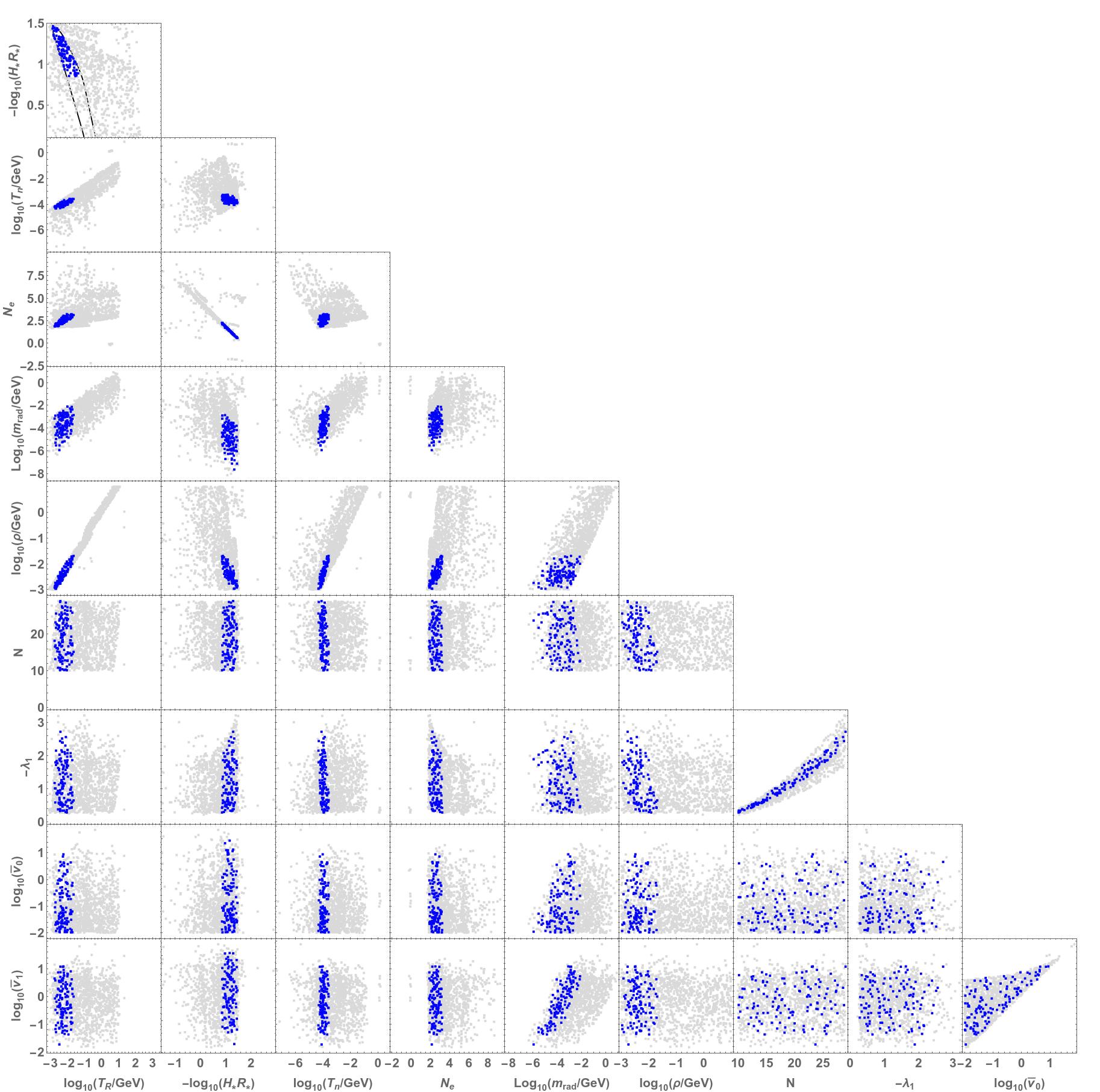} 
\caption{\it As in Fig.~\ref{fig:collision} but for the ``sound wave only" hypothesis. The ``sound wave + SMBHB" hypothesis is omitted as its 95\% C.L. favorite region practically covers the whole, considered parameter region. In the top panel, the grey points enclosed by the black line are not in blue because they fall outside the NANOGrav 95\% C.L.~region in the plane $\{\alpha_*,T_R\}$ and $\{\alpha_*,H_*R_*\}$, which reduces to $-1.5<\log_{10}(H_* R_*)< -0.85$ and $-2.8<\log_{10}(T_* / \textrm{GeV})< -1.6$ when $\alpha\gg 10$.}
\label{fig:sound}
\end{figure*}

Specifically,
the NANOGrav Collaboration considers  the ``bubble collision" and ``sound wave" contributions with their SGWB predictions parametrized  
in terms of the thermodynamic parameters $\alpha_*$, $\beta/H_*$, $T_R$, $H_* R_* = \sqrt[3]{8\pi} /(\beta/H_*)$ with the assumption $v_w=1$~\cite{NANOGrav:2023hvm}.~\footnote{The NANOGrav analysis~\cite{NANOGrav:2023hvm} also assumes small reheating, so that the nucleation, percolation and reheating temperatures are close to each other. As the explicit $T$ dependence of the SGWB is only through the redshift, the temperature $T_*$ used in the fits of Ref.~\cite{NANOGrav:2023hvm} is equivalent to our quantity $T_R$.}~The analysis considers both the presence or absence of a SMBHB SGWB component accompanying the FOPT SGWB contribution. We report their results in the top panels of Figs.~\ref{fig:collision} and~\ref{fig:sound}. In particular, the thin-dark line in the top panel of  Fig.~\ref{fig:collision} encloses the NANOGrav  95\% C.L.~favorite region for the ``bubble collision+SMBHB" hypothesis in the parameter 
space $\{T_R, H_*R_*\}$ (after marginalizing over $\alpha_*$). The thick-dark line does the same for the ``bubble collision only" hypothesis, and ``sound wave only" hypothesis in Fig.~\ref{fig:collision} and Fig.~\ref{fig:sound}, respectively. We do not consider the NANOGrav 95\% C.L.~region for the ``sound wave+SMBHB" as it practically provides no constraint (see Fig.~8 of Ref.~\cite{NANOGrav:2023hvm}). For the sake of readability, we do not display the 95\% C.L. regions in the planes $\{\alpha_*,T_R\}$ and $\{\alpha_*,H_*R_*\}$, but we impose their fulfillment. In the regime relevant in our analysis, namely  $\alpha_\ast\gg 10$, this corresponds to request $-1.5<\log_{10}(H_* R_*)< -0.85$ and $-2.8<\log_{10}(T_*/\textrm{GeV})< -1.6$ for the ``sound wave only" hypothesis, and no further constraint for the ``bubble collision+SMBHB" or ``bubble collision only" hypotheses~\cite{NANOGrav:2023hvm}. 

\begin{figure*}[t]
 \includegraphics[width=1\textwidth]{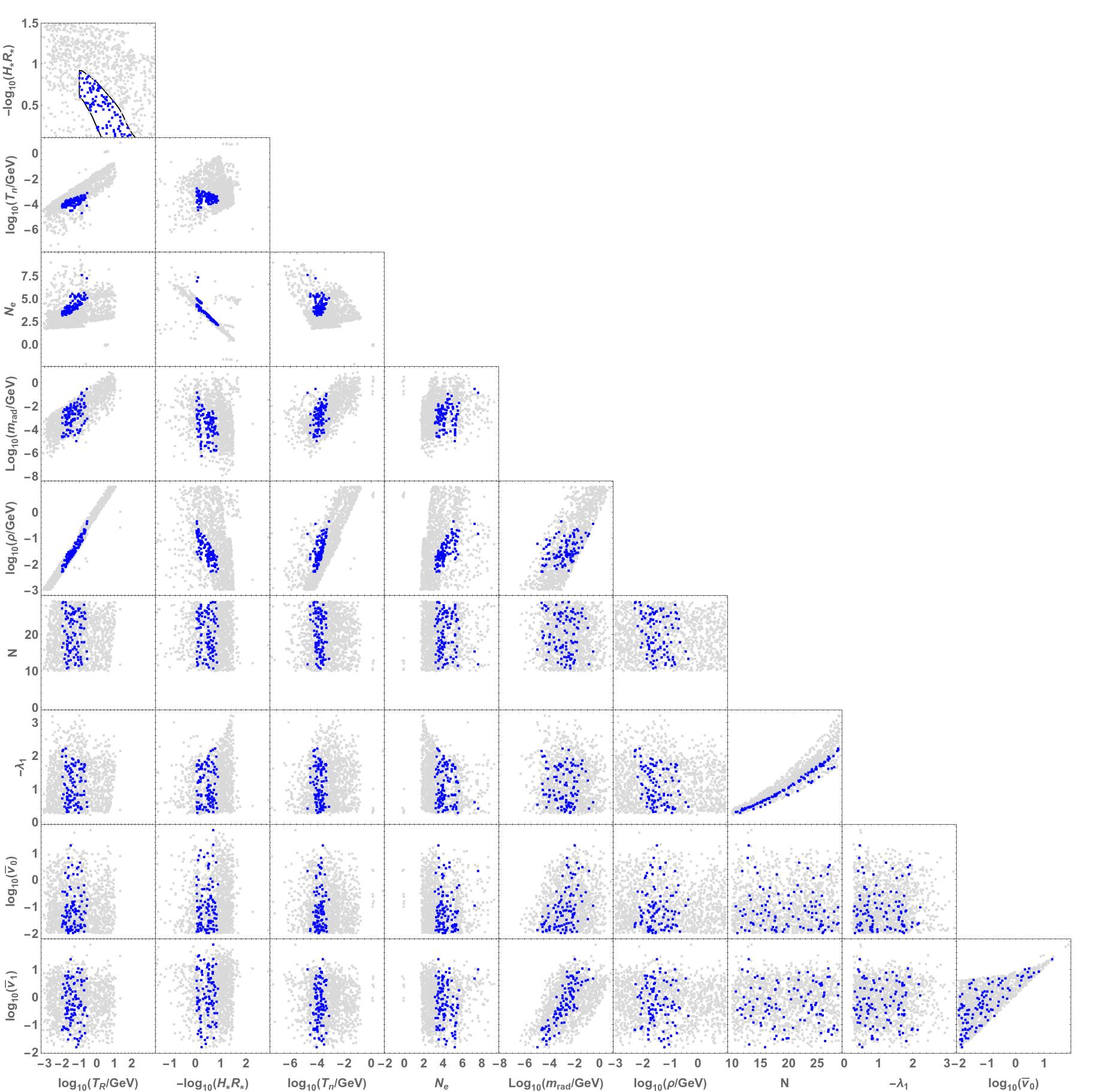} 
\caption{\it As in Fig.~\ref{fig:sound} but for the ``hydrodynamic turbulence only" hypothesis.}
\label{fig:turbulence}
\end{figure*}

The analysis in terms of the ``hydrodynamic turbulence" contribution  is carried out by the European PTA and Indian PTA Collaborations~\cite{Antoniadis:2023xlr}. They consider the `hydrodynamic turbulence" SGWB expressed  as a function of the thermodynamic parameters $\beta/H_*$, $T_R$ and an effective parameter $\Omega_*$ encapsulating the spectrum dependence on $\alpha_*$, $v_w$ and other quantities. They do not consider the case of coexistence of  ``hydrodynamic turbulence" SGWB and SMBHB SGWB. The resulting 95\% C.L. favourite region in the plane $\{T_R, H_*R_*\}$ (after marginalization over $\alpha_*$) corresponds to the area within the thick line in the top panel of  Fig.~\ref{fig:turbulence}. Being in this region is sufficient to guarantee falling inside  the 95\% C.L.~volume, as the 95\% C.L.~regions in the planes $\{\Omega_*, H_*R_*\}$ and $\{\Omega_*, T_R\}$ (marginalized over $T_R$ and $H_* R_*$, respectively) do not impose further constraints at $\log_{10}(\Omega_*)\gtrsim 0.2$, which is in the regime which our analysis focuses on (such value is reproduced for $\alpha_*\gg 10$, $v_w\simeq 1$ and large turbulence efficiency in GW production). 

To determine whether, and in case in which parameter region, the radion FOPT provides a fundamental explanation of the PTA SGWB signal, we perform a scan on the parameter space of the model.  Specifically, we explore the parameter region $N \in [10,30]$, $\log_{10}(\rho/{\rm GeV})\in [-3, 1]$, $-\lambda_1 \in [0.3,2.7]$, $\log_{10}(\bar v_{0,1})\in[-2,2]$ with flat priors. For concreteness, we fix $g_c^{\rm eff}(T)=g_d^{\rm eff}(T)\approx g_{\rm SM}^{\rm eff}(T)$. As previously stressed, our results are stable under (reasonable) variations of $g_c^{\rm eff}$ and $g_d^{\rm eff}$ (cf.~Eqs.~\eqref{eq:Ti} and \eqref{eq:TR}). Order-of-magnitude variations of $\bar v_0$ and $\bar v_1$ have no large impact of the FOPT thermodynamic parameters either.
A strong dependence on $\bar v_{0,1}$ instead appears in the ratio $m_{\textrm{rad}}/\rho$, and it is approximately given by  
$ 
m_{\textrm{rad}}/\rho \approx (\bar v_1/15) \log\left( \bar v_1/\bar v_0 \right)
$. Moreover, the parameter $u$ of the model turns out to have a dependence with $\bar v_{0,1}$ which is given by $u = \frac{1}{\bar r_1^0} \log\left( \bar v_1 / \bar v_0 \right)$ with $\bar r_1^0 \approx 40$. Then $m_{\textrm{rad}}/\rho \approx 2.7 \bar v_1 u$, so that the radion mass vanishes in the limit $u \to 0$. Notice that we have considered $\bar v_1 > \bar v_0$ in the parameter region to ensure that $m_{\textrm{rad}}^2 > 0$ and $u > 0$. For each point of the scan, we determine the FOPT thermodynamic parameters. Figures~\ref{fig:collision},  \ref{fig:sound} and \ref{fig:turbulence} summarize the outcome of the analysis. The gray [red] \{blue\} points show the results for whole sample [the fraction of the sample in the 95\% C.L.~region of the FOPT+SMBHB hypothesis] \{the fraction of the sample in the 95\% C.L.~region of the FOPT-only hypothesis\}. %Below every blue point, there is a red and a gray point, and below every red point there is a gray point.
Every blue point overlaps a red and a gray point, and every red point overlaps a gray point. We do not report on $\alpha_*$ as it always turns out to be much larger than 10 (the FOPT phenomenology is insensitive to variations of $\alpha_*$ at $\alpha_* \gtrsim 10$).  We do not calculate $v_w$ but, as argued in Ref.~\cite{Megias:2020vek},   it is reasonable to expect $v_w\simeq 1$ in a radion FOPT with $\alpha_*\gg 10$. Neither we determine the efficiency factor suppressing the GW production from turbulence, as its determination is unknown in much simpler scenarios than the radion FOPT. We postulate it to be maximal for simplicity, and leave for future studies the inclusion of this extra parameter in our analysis.

Figures~\ref{fig:collision},  \ref{fig:sound} and \ref{fig:turbulence} carry a wealth of information. The energy scale of the brane $\mathcal{B}_1$, represented by $\rho$, must be in the MeV-GeV range. Such a scale is even smaller if the sound-wave regime applies. This may constitute a challenge for some models built upon the generic RS setup here investigated (we remind that our FOPT results are practically independent of $\mathcal{L}_{\rm bulk}$ and $\mathcal{L}_{\rm a}$). In fact, it typically turns out $\rho \sim T_R \sim (10^{2}- 10^{3})T_n$ and, on top of this, some e-folds of inflation before the FOPT are unavoidable. This requires some caution to guarantee the SM sector to reach to the standard-cosmology conditions before BBN. We also highlight that there seems to be slight preference (i.e.~higher density) for low $N$ and large $\lambda_1$, but different priors should be tested before clarifying this aspect.

\section{Possible embeddings and their phenomenological signatures}
\label{sec:embeddings}
\noindent
As previously explained, the thermodynamic characteristics of the radion FOPT are almost independent of the particular embedding, namely $\mathcal{L}_{\rm bulk}$ and $\mathcal{L}_{a}$. The phenomenology instead strongly depends on it. Before identifying some promising embedding options, we summarize some key results and ideas useful for our purpose. Given that the scale of the IR brane $\mathcal B_1$ is at the MeV-GeV scale, the bulk Lagrangian $\mathcal{L}_{\rm bulk}$ cannot contain any SM fields as their KK resonance masses would be too low and should have already been discovered at LHC and through electroweak precision measurements. In other words we will concentrate the SM in some 4D brane with cutoff higher than the TeV scale. Two simple options appear at this level.
\begin{enumerate}
\item
The simplest option is having a model with two branes ($n=2$), the UV brane $\mathcal B_0$ (at the Planck scale) and the IR brane $\mathcal B_1$ (at the sub-GeV scale). The SM should be localized in the UV brane, with Planckian cutoff, so that the running of SM couplings is logarithmic and unification of gauge couplings can take place at high scales provided the appropriate matter is added to the SM. In this model the warp factor does not solve the hierarchy problem, but the role of the warping is precisely to create the scale of the IR brane for which the SGWB is eventually being detected at PTA Collaborations. The hierarchy problem could be mitigated e.g.~if the SM in the UV brane is completed by supersymmetry, in which case we will have the bonus of gauge coupling unification at scales $\sim 10^{16}$ GeV. In the brane $\mathcal B_1$ only some hypothetical dark sector can eventually live. Only the graviton (and radion) sectors are propagating in the bulk of the extra dimensions with KK modes at the sub-GeV scale.
\item
An alternative option is a model with three branes ($n=3$), the UV brane $\mathcal B_0$ (at the Planck scale), the most IR brane $\mathcal B_1$ (at the sub-GeV scale) and an intermediate brane $\mathcal B_T$ (at the TeV scale)~\footnote{For notational convenience we are renaming the intermediate brane $\mathcal B_I$ at the TeV scale in this section as $\mathcal B_T$.}. Models with multiple intermediate branes have been studied e.g.~in Refs.~\cite{Lykken:1999nb,Lee:2021wau}. The three branes are located at $r_0\ll r_T\ll r_1$, or using conformal coordinates $k z\simeq e^{k r}$, at $z_0\ll z_T\ll z_1$, such that $z_0k=z_T\rho_T=z_1\rho=1$, where $\rho_T\simeq $ TeV. In this model the SM is located in the $\mathcal B_T$ brane, so that the hierarchy problem is solved by the warp factor. This model is an IR extension of the RS model. The graviton is propagating in the whole bulk providing graviton KK modes at the sub-GeV scale. There are two radions: the radion $r_T(x)$ which stabilizes the $\mathcal B_T$ brane at $\langle r_T\rangle=\rho_T$, at temperatures $T\sim \rho_T$, with support in the region $[r_0,r_T]$ and KK masses in the sub-TeV scale; and the radion $r(x)$, which stabilizes the $\mathcal B_1$ brane with respect to $\mathcal B_T$ at $\langle r\rangle=\rho$ at temperatures $T\sim \rho$, with support in the region $[r_T,r_1]$ and KK masses in the sub-GeV scale. For hierarchically related phase transitions, i.e.~for $\rho_T\gg \rho$, we can assume that, at temperatures $T\gtrsim \rho$, the first phase transition has already settled the heavy radion $r_T(x)$ in its vacuum expectation value $\langle r_T\rangle=\rho_T$, while keeping the light radion in the symmetric phase $\langle r\rangle=0$, just leaving a vacuum energy, so that the second transition may proceed without any influence of the heavy radion, which has been integrated out. We then assume a two-step phase transition.
The brane $\mathcal B_1$ is then stabilized by the radion $r(x)$ leaving the SGWB detected by the PTA Collaborations.
\end{enumerate}

\subsection{The graviton sector}
\noindent
The model has a graviton sector, which propagates in the whole bulk, with a massless zero mode and an infinite number of resonances with masses $m_n$ such that
$m_1\simeq 3.9 \rho \dots$. It generates a deviation of the Newtonian potential given by~\cite{Fichet:2022ixi,Fichet:2022xol}
\begin{equation}
V(R)=-\frac{m_1 m_2}{8 \pi M_{\rm Pl}^2} \frac{1}{R} (1+\Delta(R)) \,,
\end{equation}
where
\begin{equation}
\Delta(R)=\frac{4}{3}\sum_{n=1}^\infty e^{-m_n R}\simeq \frac{4}{3}e^{-m_1 R} \,,
\end{equation}
which is dominated by the first KK mode. The experimental constraints on Yukawa-type deviations from Newtonian gravity yield for KK gravitons 
$1/m_1 \lesssim 30 \, \mu m$~\cite{Smullin:2005iv,Geraci:2008hb,Adelberger:2022sve}, or 
%(using the conversion $1\,\textrm{cm}^{-1}=1.97\cdot 10^{-14}$ GeV)
%
\begin{equation}
m_1\gtrsim 10^{-11}\ \textrm{GeV}  \,.
\end{equation}
This is a very mild constraint as a consequence of the exponential suppression of the correction. 

The graviton coupling to matter in the brane $\mathcal B_b$ is given by
\begin{equation}
\mathcal L=-\frac{1}{M_5^{3/2}}h_{\mu\nu}(z_b,x)T^{\mu\nu}(x) \,.
\end{equation}
The equations of motion of the graviton sector can be found e.g.~in Ref.~\cite{Davoudiasl:1999jd}. There is a massless zero mode, the physical graviton, while the mass spectrum of the KK modes is given by $m_n=x_n\rho$, where $x_n$ is determined by the condition $J_1(x_n)=0$. As for the wave function of KK modes, after making the expansion in modes $h_{\mu\nu}=\sum_n h^{(n)}(z) h_{\mu\nu}^{(n)}(x)$ one easily finds 
\begin{equation}
h^{(n)}(z)=\sqrt{k}\frac{(kz)^2}{kz_1}\, \frac{J_2(m_nz)}{J_2(m_n z_1)} \,.
\end{equation}
 The coupling of KK modes to the $\mathcal B_b$ brane is then written as
 \begin{equation} 
 \mathcal L=-\sum_n c_n(z_b) h_{\mu\nu}^{(n)}(x) T_{\mu\nu}(x) \,,\qquad c_n(z_b)=\frac{1}{M_{\rm Pl}}\frac{(k z_b)^2}{kz_1}\frac{J_2(x_n z_b/z_1)}{J_2(x_n)} \,.
 \end{equation}
 As for the possible three branes we can have in the above models, the coupling is:
 \begin{itemize}
 \item
 For $z=z_1$ we have
 \begin{equation}
 c_n(z_1)=\frac{k}{M_{\rm Pl}}z_1=\frac{k}{M_{\rm Pl}}\, \frac{1}{\rho} \,.
 \end{equation}
 This coupling is very mildly suppressed for the present models where $\rho$ is at the sub-GeV scale. In RS-type models where the SM is localized at the $\mathcal B_1$ brane and $\rho$ is at the TeV scale, this reproduces the usual result that the KK modes are coupled with the strength $\sim 1/$TeV.
 In our models we only assume that a dark sector could be localized on the $\mathcal B_1$ brane.
 \item
 For $z=z_T$ we have
 \begin{equation}
 c_n(z_T)=\frac{k}{M_{\rm Pl}}\frac{z_T^2}{z_1}\frac{J_2(x_n z_T/z_1)}{J_2(x_n)}\simeq \frac{k}{0.1M_{\rm Pl}} z_T \epsilon_n(z_T) \,,  \qquad
 \epsilon_n(z_T)=0.1\left( \frac{z_T}{z_1} \right)^3\frac{x_n^2}{8J_2(x_n)} \,,
 \label{eq:cn}
 \end{equation}
 where we have normalized the ratio $k/M_{\rm Pl}$ to 0.1, and made use of the property $J_2(x)=x^2/8+\mathcal O(x^4)$ valid for $x\ll 1$. The prefactor of $\epsilon_n(z_T)$ gives the usual 1/TeV coupling of KK gravitons in RS models, while $\epsilon_n(z_T)$ is an extra suppression factor. 
 For instance, for the first KK mode we have that $x_1\simeq 3.9$ and $J_2(x_1)\simeq 0.4$ so that the suppression factor, for $\rho_T\gtrsim 1$ TeV and $\rho\lesssim 1$ GeV, is given by $\epsilon_1(z_T)\lesssim 5\cdot 10^{-10}$. This tiny coupling makes it possible to evade present ATLAS bounds of $m_1\gtrsim 2$ TeV~\cite{ATLAS:2014pcp}, valid for $\epsilon_1(z_T)=1$.
 \item
 For $z=z_0$ the suppression factor stems from Eq.~(\ref{eq:cn}) by replacing $z_T\to z_0$. It is so tiny ($\epsilon_n(z_0)\sim 10^{-55}$)
that in practice graviton KK modes are decoupled from the SM, when the latter is localized in the UV brane $\mathcal B_0$, as in model 1) above. 
 \end{itemize}
 
\subsection{The radion sector}
\noindent
The radion corresponds to scalar perturbations $F(r,x)$ of the metric as $ds^2=e^{-2(A+F)}\eta_{\mu\nu}-(1+2F)^2 dr^2$. Its coupling to matter localized on the $\mathcal B_b$ brane is given by
\begin{equation}
\mathcal L=-F(z_b,x)T^\mu_\mu(x) \,.
\end{equation}
In the limit of small backreaction the relation between the canonically normalized 4D radion $r(x)$, which stabilizes the $\mathcal B_1$ brane, and the metric perturbation $F(z,x)$ is given by~\cite{Csaki:2007ns}
\begin{equation}
F(z,x)=\frac{1}{\sqrt{6}}\frac{z_b^2}{z_1 z_T} z_T r(x)=\epsilon(z_b) \frac{r(x)}{\Lambda_r} \,,\qquad \epsilon(z_b)=\frac{z_b^2}{z_1 z_T} \,,
\end{equation}
where $\Lambda_r=\sqrt{6}\rho_T$ provides the radion coupling strength in the RS model, and $\epsilon(z_b)$ is an extra suppression factor.

The mass of the radion is given by $m^2_{\rm rad}=\rho^2/\Pi_{\rm rad}$. The function $\Pi_{\rm rad}$ is a cumbersome expression with $d \Pi_{\rm rad}/d\lambda_1 > 0$ and $\Pi_{\rm rad}(\lambda_1)\sim 10^{2}-10^{4}$ for $\lambda_1 \in [-10,-0.1]$, but a simple phenomenological good enough approximation is provided in Sec.~\ref{sec:FOPT}. We refer the reader to Eq.~(31) in Ref.~\cite{Megias:2020vek} for the full expression that we adopt in our analysis.
\begin{itemize}
\item
For $z_b=z_1$ we can see that the suppression gets reduced from the typical RS result, as the coupling of the radion has the strength $1/\rho$. This coupling would only apply to a hypothetical dark sector localized on the $\mathcal B_1$ brane.
\item
For $z_b=z_T$ we can see that the suppression factor is given by
\begin{equation}
\epsilon(z_T)=\frac{z_T}{z_1}=\frac{\rho}{\rho_T}\simeq 10^{-3} \,.
\end{equation}

This radion decays into the SM matter in model 2) above, for which the SM is localized in the $\mathcal B_T$ brane. In particular its coupling to two gluons $gg$ (leading to the main decay channel) is governed by the interaction Lagrangian given by~\cite{Giudice:2000av,Csaki:2007ns}
\begin{equation}
\mathcal L=\epsilon(z_T)c\,\frac{\alpha_{\rm s}}{8\pi}\frac{r(x)}{\Lambda_r}G_{\mu\nu}G^{\mu\nu} \,, \qquad c = 7+4\,\frac{1}{2}\,\frac{4}{3}=
\frac{29}{3}\,,
%
%\frac{7}{3}+7-\frac{16}{9} = \frac{68}{9} \,,
\end{equation}
where we have included the contribution from the localized trace anomaly (first term in $c$), and from triangular diagrams involving quarks $Q$ such that $4 m_Q^2\gg m_r^2$ (second term in $c$), considering the second and third generation quarks.
The decay width, and lifetime, into $gg$ is given by
\begin{equation}
\Gamma(r\to gg)=\frac{c^2 \epsilon^2(z_T)\alpha_{\rm s}^2}{32\pi^3} \frac{m_r^3}{\Lambda_r^2}\simeq 2\cdot 10^{-19} \left( \frac{\rho}{\rm GeV}\right)^5
\textrm{ GeV}\,, \qquad \tau(r\to gg)\simeq 3\cdot 10^{-6}\left( \frac{\rm GeV}{\rho}\right)^5 \, \textrm{s}%\simeq  10^4 \, \textrm{m}
\end{equation}
where, in the last equality, we have used the numerical values $\Lambda_r=2.5$ TeV, $m_r=0.1\rho$, and normalized $\rho$ to $1$~GeV. Investigating the phenomenology and cosmology of such light feebly coupled radion is beyond the scope of the present paper and should be done in a separate publication.

\item
Finally for $z_b=z_0$, which is relevant for model 1) above, the coupling of the radion to the SM, localized in the $\mathcal B_0$ brane, has a huge suppression factor given by $\epsilon(z_0)=z_0^2/(z_1 z_T)$, and taking typical values $\rho\simeq 1$ GeV, $\rho_T\simeq 1$ TeV we get $\epsilon(z_0)\simeq 10^{-33}$ and the previous calculation would lead to a lifetime of the radion for the $\gamma\gamma$ channel larger than the present age of the universe. This particle is then stable in cosmological times and could play the role of dark matter provided its abundance can be conveniently controlled. However if there is some dark sector in the $\mathcal B_1$ brane (containing e.g.~dark matter) it could easily decay into it. Again, constructing a complete model with a dark sector is not going to affect appreciably the FOPT and is outside the scope of the present paper. 
\end{itemize}

As we aim at addressing the PTA observation, we do not investigate extensions of this proposal. Adding new light fields in the $\mathcal B_1$ brane may turn a viable option to address the Dark Matter puzzle, and many of the 4D extensions of the SM investigated in the literature can be accommodated in the UV brane. Most of such extensions are not expected to qualitatively change our main conclusions obtained in the minimal scenario.

\section{Conclusions}
\label{sec:conclusions}
\noindent
Recent results on the detection of a SGWB in the nHz frequency band by the pulsar timing array collaborations lead, as a possible BSM explanation, to the existence of a strongly-first order (dark) phase transition with scale in the GeV range. We have first proposed in this paper a modelization of such phase transition in a simple warped five-dimensional model with a UV brane, at the Planck scale, and a dark brane at the GeV scale. The phase transition is the confinement/deconfinement phase transition where the radion field fixing the inter brane distance undergoes a transition from its deconfined state (above the nucleation temperature) in the 5D BH configuration, to its confined state (below the nucleation temperature) in the thermal 5D warped space. The model can be considered as a discretization of the RS2 model~\cite{Randall:1999vf} by means of the introduction of a brane at the GeV scale. This model cannot support SM fields propagating in the bulk of the extra dimension, so that the SM (as in the RS2 model) has to be localized in the UV brane. On the other hand the dark brane can contain a dark sector, which the radion field interact with, and might contain dark matter candidates. The details of the matter content of the dark sector are not relevant for the present paper (as soon as they interact with the SM only through gravitational interactions) so we do not specify them here. As it is clear from its conception, as the SM is localized on the UV brane, the model does not solve the hierarchy problem (whose solution would more conventionally require the introduction of supersymmetry in the UV brane) but just introduces the dark brane to make contact with the recent PTA results.

Furthermore, models with multiple hierarchies~\cite{Lee:2021wau} have been studied in the literature, and can be equally used to describe the PTA data along with solving the hierarchy problem. For instance between the UV and dark brane, one could introduce a TeV brane, where the SM can be localized, as in the original RS framework. In that case there should be two strongly-first order phase transitions induced by the two radion excitations (describing the fluctuations of the TeV and dark branes, respectively): \textit{i)} a heavy sub-TeV radion, with support in the region between the UV and TeV scales; and, \textit{ii)} a light sub-GeV radion, with support in the region between the TeV and dark branes. As gravitons propagate along the whole extra dimension, their KK modes are at the GeV scale. The first phase transition, driven by the heavy radion $r_T(x)$, should happen at temperatures of the TeV and reproduces the confinement/deconfinement phase transition studied in RS models~\cite{Creminelli:2001th}. This FOPT should generate a SGWB that should be detected at LISA and ET~\cite{Megias:2018sxv}. 
The second confinement/deconfinement phase transition, driven by the light radion $r(x)$, should happen at temperatures of the order of the GeV. The latter phase transition, that has been analyzed throughout this paper, is straightforward as we can integrate out the GeV (or heavier) KK modes and remain with the SM and the light radion as the only degrees of freedom. However for the former one, as the heavy radion is normally heavier than the graviton KK modes, has to happen in the presence of the irreducible background of the graviton KK modes. They can affect the phase transition in a way that has to be thoroughly taken into account. The details of this analysis are postponed for future investigations.  

%

%\bibliography{refs}

\begin{acknowledgments}
  \noindent
  
The work of EM is supported by the project PID2020-114767GB-I00
funded by MCIN/AEI/10.13039/501100011033, by the FEDER/Junta de
Andaluc\'{\i}a-Consejer\'{\i}a de Econom\'{\i}a y Conocimiento 2014-2020
Operational Programme under Grant A-FQM-178-UGR18, and by Junta de
Andaluc\'{\i}a under Grant FQM-225. The research of EM is also supported
by the Ram\'on y Cajal Program of the Spanish MICIN under Grant RYC-2016-20678. GN is partly supported by the
grant Project. No. 302640 funded by the Research Council of Norway. 
The work of MQ is supported by the Departament d'Empresa i Coneixement, Generalitat de Catalunya, Grant No.~2021 SGR 00649, and by the Ministerio de Econom\'{\i}a y Competitividad, Grant No.~PID2020-115845GB-I00. IFAE is partially funded by the CERCA program of the Generalitat de Catalunya.
\end{acknowledgments}

\bibliography{refs_rev}

\end{document}